\documentclass[conference]{IEEEtran}
\IEEEoverridecommandlockouts
\usepackage[utf8x]{inputenc}
\usepackage[table,xcdraw]{xcolor}
\usepackage{graphicx}
\usepackage[pdfa]{hyperref}
\usepackage{balance}
\usepackage{cite}
\usepackage{csquotes}
\usepackage[english]{babel}
\usepackage{amsmath,amssymb,amsfonts}
\usepackage{algorithmic}
\usepackage{graphicx}
\usepackage{textcomp}
\usepackage{cleveref}
\usepackage{xcolor}
\def\BibTeX{{\rm B\kern-.05em{\sc i\kern-.025em b}\kern-.08em
    T\kern-.1667em\lower.7ex\hbox{E}\kern-.125emX}}
\begin{document}

\title{Multimodal music information processing and retrieval: survey and future challenges}

\author{
    \IEEEauthorblockN{
        Federico Simonetta,
        Stavros Ntalampiras,
        Federico Avanzini
}
    \IEEEauthorblockA{
        LIM -- Music Informatics Laboratory\\
        Department of Computer Science\\
        University of Milan\\
        Email: \{name.surname\}@unimi.it    }
}

\maketitle

\begin{abstract} Towards improving the performance in various music information
    processing tasks, recent studies exploit different modalities able to
    capture diverse aspects of music.  Such modalities include audio
    recordings, symbolic music scores, mid-level representations, motion and
    gestural data, video recordings, editorial or cultural tags, lyrics and
    album cover arts.  This paper critically reviews the various approaches
    adopted in Music Information Processing and Retrieval, and highlights how
    multimodal algorithms can help Music Computing applications.  First, we
    categorize the related literature based on the application they address.
    Subsequently, we analyze existing information fusion approaches, and we
    conclude with the set of challenges that Music Information Retrieval and
    Sound and Music Computing research communities should focus in the next
    years.

\end{abstract}


\section{Introduction}\label{sec:intro}

    Beginning with the oldest evidence of music notation, music has been described
    in several different forms\cite{Bent2001}. Such descriptions have been used by
    computational systems for facilitating music information computing tasks.
    Interestingly, when observing the history of music, one can see how the various
    descriptive forms have gradually emerged with a strict dependence both on
    technology advancements and changes in music practices.

    Initially, no written description systems for music existed besides text.
    Between the 6th-7th cen., Isidore of Seville, Archbishop and theologian, wrote
    that no melody could be written. Indeed, the first systems to memorize music were
    based solely on lyrics and only later some signs over the words appeared.
    Such notation, called \textit{neumatic}, evolved in more complex forms,
    which differed from region to region. Due to the need of more powerful
    tools to express music features, new notation systems, called \textit{pitch
    specific}, took place, such as the \textit{alphabetic} and the
    \textit{staff}-based notations. In particular, the system introduced by
    Guido d'Arezzo (10th-11th cen.) was particularly successful and similar
    conventions spread all over Europe.  Music notation was now able to
    represent text, pitches and durations at the same time. During the
    following centuries, other types of symbols were introduced addressing
    directly the performer towards peculiar colors, or sentiment(s). At the
    crossing of the 16th and 17th cen., Opera was born in Italy,
    after a long tradition of plays, including Greek drama, medieval
    entertainers and renaissance popular plays (both liturgic and
    profane)\cite{Brown2001}. The tremendous success of the Opera in Italy and
    then in the rest of Europe, determined a fundamental way to connect music
    and visual arts for the future centuries. A turning point in the history of
    music description systems was the invention of the \textit{phonograph
    cylinder} by Thomas Edison in 1877 and the \textit{disc phonograph}
    diffused by Emile Berliner ten years later\cite{Mumma2003}. In the same
    years, Edison and the Lumi\`ere brothers invented the first devices to record
    video\cite{Cook2018}. Since then, a number of technologies were born paving
    the way for new music description systems. With the invention of computers
    and the beginning of the digital era, the elaboration of sound signals
    highlighted the need for more abstract information characterizing audio
    recordings. Thus, researchers started proposing \textit{mid-level}
    representations\cite{Kitahara2010}, with reference to \textit{symbolic} and
    \textit{physical} levels\cite{Vinet2004}.  Nowadays, the
    availability of vast, easily accessible quantities of data, along with appropriate modern
    computational technologies, encourages the collection of various types of
    \textit{meta-data}, which can be either \textit{cultural} or
    \textit{editorial}\cite{Pachet2005}.

    From a cognitive point of view, the connecting, almost evolutionary, element
    between the above-mentioned representations is that each one relates to a
    different abstraction level. Psychology, indeed, is almost unanimous in
    identifying an abstraction process in our music cognition\cite{Deutsch2013a}:
    we can recognize music played on different instruments, with different timings,
    intensity changes, various metronome markings, tonalities, tunings, background
    noises and so on. The different descriptions of music developed in different
    era or contexts, can be seen as an answer to the necessity of representing new
    modalities --~such as the visual one~-- or new unrevealed abstraction levels --~such as the audio recordings and the mid-symbolic levels, or the pitch specific
    notation compared to the neumatic one.

    Aside from these historical and cognitive considerations, it is a fact that in
    the last two decades researchers have obtained better results through
    multimodal approaches in respect to single-modalities approaches\cite{Atrey2010, Baltrusaitis2018}. As Minsky
    said\cite{Minsky1991}: 

    \begin{displayquote} 
        \textit{
            To solve really hard problems, we'll have to use several different
            representations.  
    }
    \end{displayquote}

    We argue that music processing tasks can benefit profoundly from multimodal
    approaches, and that a greater focus is needed by the research community in
    creating such a synergistic framework. A fundamental step would be the study
    and design of suitable algorithms through which different modalities can
    collaborate. Then, a particular effort should be devoted in developing the
    needed technologies. In fact, given the course of history summarized above, we
    could expect that in the
    future, new disparate music representations will be born.

    In this paper, we review the existing literature about Music Information
    Retrieval techniques which exploit multiple descriptions of
    music to the end of \textit{multimodal fusion}\cite{Essid2012}.
    The paper is organized as follows: in \cref{sec:overview}, we give some basic definition and discuss previous reviews on similar topics to explain the
    categorization and the taxonomy we used.
    \Crefrange{sec:tasks}{sec:fusion} describe the different tasks 
    faced with multimodal approaches, the various
    features extracted, the preprocessing steps and the fusion approaches adopted
    in literature; in \cref{sec:future} we express our idea about how the multimodal
    paradigm can be enhanced.

\section{Definitions, taxonomy and previous reviews}\label{sec:overview}

    We have found no univocal definition of modality. In the music computing literature, authors use the
    word
    \textit{multimodal} in two main contexts:
    \begin{itemize}
        \item in computational psychology, where \textit{modality} refers to a human sensory channel;
        \item in music information retrieval, where \textit{modality} usually refers a source of
        music information;
    \end{itemize}
    Since we are focusing on music information retrieval methods, to the purpose of the present
    paper, with \textit{modality} we mean a specific way to digitize music information. 
    Different modalities are obtained through different transducers, in different places or times, and/or belong to different media. Examples of modalities that may be associated to a single piece of music include audio, lyrics, symbolic scores, album covers, and so on.
     
    Having defined what we mean by modality, we define \textit{multimodal music information processing} as 
    an MIR\cite{Schedl2014} approach  
    which takes as input multiple modalities of the same piece of music.
    All the papers which we are going to discuss show methods which take as input various music
    representations. Conversely, we are not
    considering those approaches which exploit features derived through different methods from the
    same modality: an example is pitch, rhythmic and timbral features, when they are all derived from the audio\cite{Tzanetakis2002}. Similarly we are not considering approaches which process multiple
    representations of the same modality: an example is spectrograms (treated as 2D images) and traditional time-domain acoustic features\cite{Nanni2016}, which are both derived from the audio.
    Moreover, we do not focus on general multimodal sound
    processing: the idea which moves our effort is that music is characterized by the
    \textit{organization} of sounds in time; thus, we are interested in exploiting this
    organization, which is not available in general sound processing.

    One previous review on multimodal music processing was written in
    2012\cite{Essid2012}. However, that work was more focused on a 
    few case studies rather than on an extensive survey. The authors recognized
    a distinction between
    ``the effort of characterizing the \textit{relationships} between the
    different modalities'', which they name \textit{cross-modal processing}, and
    ``the problem of efficiently combining the information conveyed by the
    different modalities'', named \textit{multimodal fusion}. To our analysis,
    this distinction is useful if with \textit{cross-modal processing} we mean
    the end-user systems which offer an augmented listening experience
    by providing the user with additional information. If this is the case, we
    are primarily interested in \textit{multimodal fusion}; nevertheless, some
    synchronization algorithms, which are classified as
    \textit{cross-modal processing} by the previous authors\cite{Essid2012}, are 
    used as pre-processing steps in other works. Because of this ambiguous 
    distinction, we base our classification on the performed task rather than on
    the processing stage --~see \cref{sec:tasks}.

\begin{figure}[]
        \centerline{\includegraphics[width=0.5\textwidth]{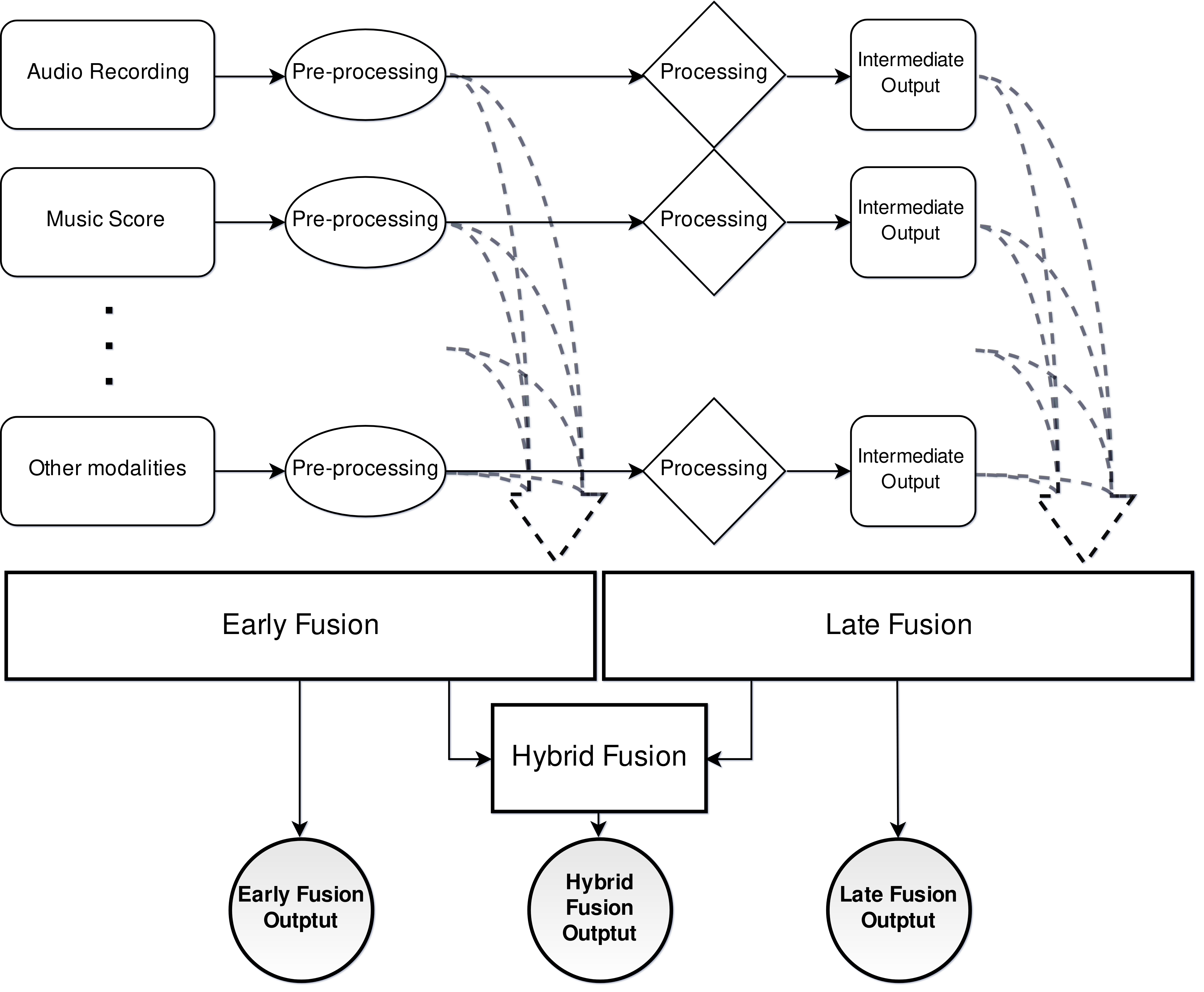}}
        \caption{
            Diagram showing the flow of information in \textit{early}-fusion
            and \textit{late}-fusion. Early fusion process takes as input the output
            of the pre-processing of the various modalities, while the late fusion
            takes as input the output of specific processing for each modality.
            \textit{Hybrid fusion}, instead, uses the output of both \textit{early} 
            and \textit{late} fusion.
        }
        \label{fig:fusion}
    \end{figure}

    Almost all authors dealing with multimodal information fusion talk about
    two approaches: \textit{early fusion} and \textit{late fusion}. \Cref{fig:fusion}
    shows the main difference between the two approaches: in \textit{early fusion}, data 
    is used ``as is'' in one single processing algorithm which fuse the data representation,
    while in \textit{late fusion} data from each modality is first processed
    with specific algorithms and then all the output are merged, so that it is
    the output to be fused and not the data. Because of this, \textit{early
    fusion} is also called \textit{feature-level fusion}, and \textit{late
    fusion} is also called \textit{decision-level fusion}, even if features
    extraction and decision algorithms are not the only approaches for multimodal
    processing. Some reviews\cite{Atrey2010} also talk about \textit{hybrid fusion} for multimedia analysis,
    but we have found no example in the music domain.

\begin{figure*}[t]
        \centerline{\includegraphics[width=0.7\textwidth]{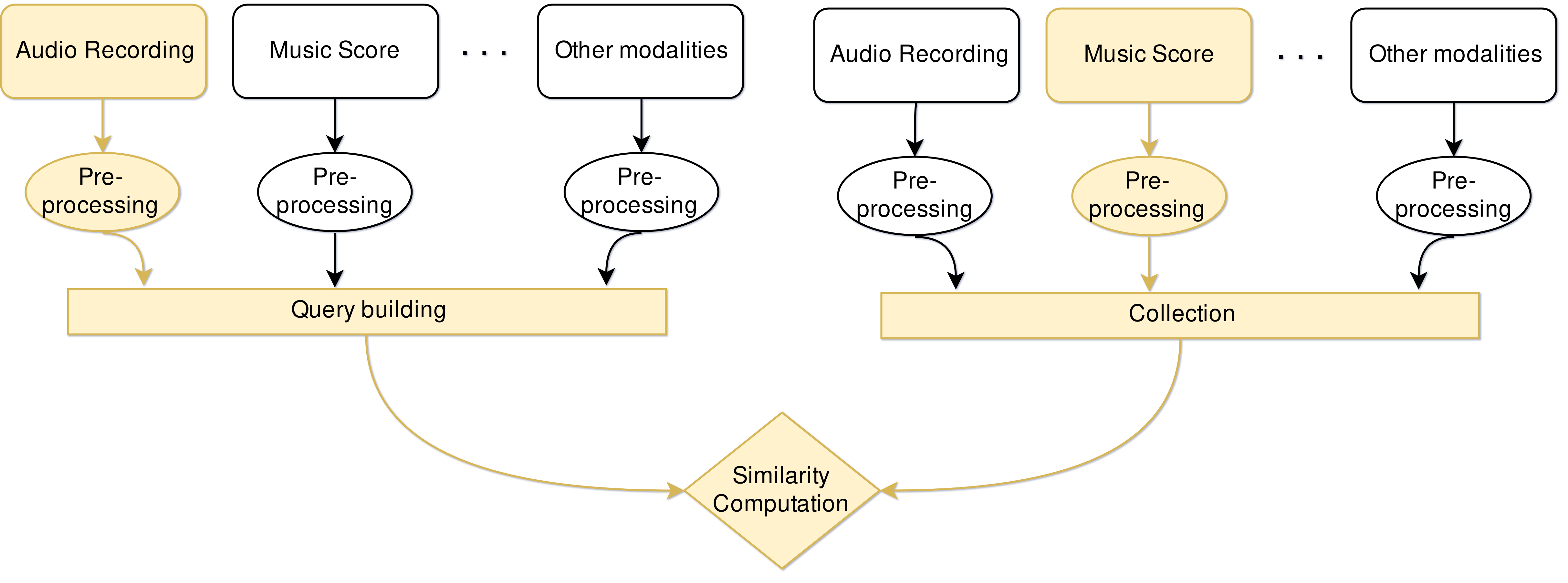}}
        \caption{Multimodal retrieval: usually, the query and the collection
        contain different modalities, so that the diagram should be collapsed to
        the highlighted elements; however a more general
        case is possible\cite{Zhonghua2013}, in which both the query and the collection
        contain multiple modalities.}
        \label{fig:retrieval}
    \end{figure*}

    Finally, we have found useful to introduce a new diagram to represent the
    data flow in retrieval systems (see \cref{fig:retrieval}). Indeed, in most of
    these systems, one modality is used to query a database for retrieving
    another modality; in such cases, no fusion exists, but just a data conversion and
    a similarity computation.

    An exhaustive and continuously updated table, which summarizes all the works reviewed in this paper, is available
    online.\footnote{\label{foot:link}Link: \href{https://frama.link/multimodal-MIR}{https://frama.link/multimodal-MIR}}

\section{Multimodal music processing tasks}\label{sec:tasks}
    To date, several tasks have been experimented in multimodal approaches. We
    found two possible categorizations for the application level:

    \begin{itemize}

        \item \textbf{less} vs \textbf{more} studied tasks: some tasks have
            been extensively studied with a multimodal approach, such as
            \textit{audio-to-score alignment}, \textit{score-informed source
            separation}, \textit{music segmentation}, \textit{emotion} or
            \textit{mood} recognition; other tasks, instead, have been little
            explored and are worth of more attention.
        
        \item \textbf{macro-task} based categorization: we identified $4$ different
            macro-tasks, that are a partial re-elaboration of a previous effort\cite{Schedl2014}:
            \textit{classification} of music, \textit{synchronization}
            of different representations, \textit{similarity} computation between two 
            or more modalities, and \textit{time-dependent representation}.

    \end{itemize}

\begin{figure*}[]
        \centerline{\includegraphics[width=0.7\textwidth]{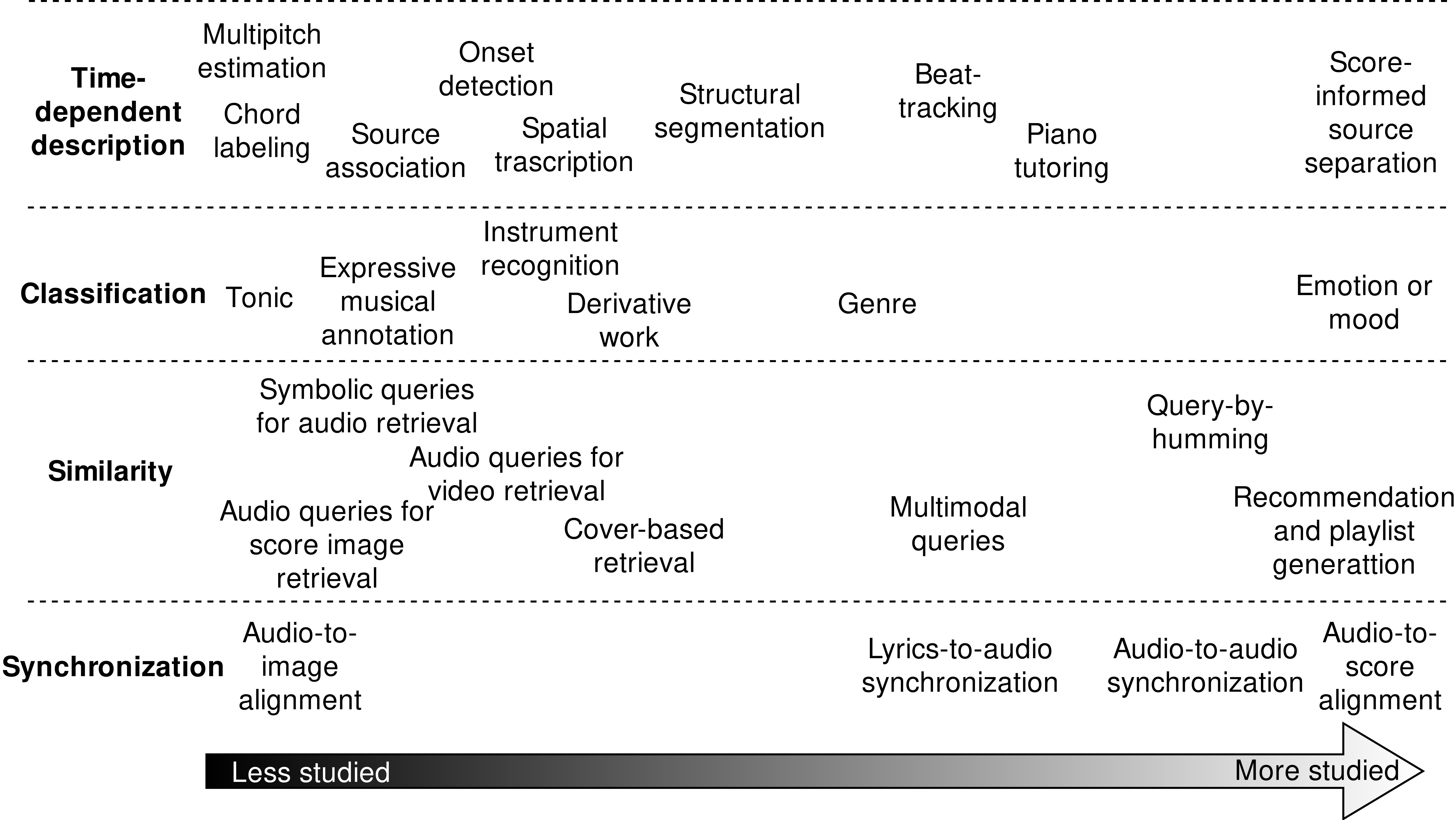}}
        \caption{The tasks identified in literature, divided in $4$ macro-tasks and
        plotted along a \textit{less} - \textit{more} studied axis. Tasks for which only one paper has been found appear at the left-side (\textit{less studied}); at the rightmost side are tasks for which extensive surveys are already available; the other tasks are placed in the remaining space proportionally to the number of corresponding papers found in literature. All references to these tasks can be found in the discussion and in the online spreadsheet -- see \cref{foot:link}. Note that labels refer to the multimodal approach at hand and not to generic MIR tasks -- e.g. \textit{genre} classification task is intended to be performed with a multimodal approach and thus it has been less studied than \textit{emotion or mood} classification in the context of multimodal approaches. }
        \label{fig:tasks}
    \end{figure*}

    \Cref{fig:tasks} outlines all the tasks that we found in the literature. Here, instead,
    we are going to briefly describe each task and how it has been fulfilled by
    exploiting a multimodal approach.

    \subsection{Synchronization}\label{sec:sync}
        Synchronization algorithms aim at aligning in time or space
        different modalities of music, i.e. creating associations between
        points in different modalities. They can be performed both in real-time
        and offline. In the real-time case, the challenge is to predict if a
        new event discovered in a real-time modality --~e.g. an onset in the
        audio~-- corresponds to an already known event in another off-line
        modality --~e.g. a new note in the score. Off-line synchronization,
        instead, is usually referred to as \textit{alignment} and involves the
        fusion of multiple modalities by definition. Well-studied alignment
        algorithms include \textit{audio-to-score} alignment\cite{Mueller2015},
        \textit{audio-to-audio} alignment\cite{Mueller2015} and
        \textit{lyrics-to-audio} alignment\cite{Fujihara2012}. An interesting task 
        is to align the audio recording to the images, without using any symbolic data
        \cite{Dorfer2017}.  Very often, alignment algorithms are a fundamental
        pre-processing step for other algorithms --~see
        \cref{sec:preprocessing}.

    \subsection{Similarity} 
        With \textit{similarity}, we mean the task of computing
        the amount of similarity between the information content of different 
        modalities. Often, this task has the purpose of retrieving documents 
        from a collection through a query, which can be explicitly expressed 
        by the user or implicitly deduced by the system. The multimodal approach,
        here, can exist either in the different modalities between the query
        and the retrieved documents or in the query itself. A very common example
        of explicit queries for retrieving
        another modality is \textit{query-by-humming} or \textit{query-by example},
        in which the query is represented by an audio recording and
        the system retrieves the correct song; this task is usually performed
        with two main approaches: by using a collection of recordings in a
        single-modality fashion, or by exploiting multimodality with a
        collection of symbolic data\cite{Kotsifakos2012, MIREX2016}. An example
        of \textit{implicit} query systems, instead, are
        recommender systems and playlist generators, where the user is
        usually not aware of which specific parameters are used for the
        recommendations; most of the recent research in this field tries to
        exploit multimodal approaches --~also called \textit{hybrid}~--
        involving \textit{metadata}, user \textit{context}, \textit{audio}
        features\cite{Bonnin2014, Deshmukh2018}. An emerging field in the
        retrieval context is the so-called \textit{multimodal queries}, where
        the user can explicitly create a query by using different parameters
        for different modalities\cite{Mueller2019, Zhonghua2013}. Following this line of
        thought, some researchers 
        devised and studied novel tasks
        in the context of multimodal music retrieval. Some example are: a
        system for retrieving music score images through audio queries\cite{Dorfer2017}; an algorithm to retrieve the cover of a
        given song\cite{Correya2018}; systems to retrieve
        audio recordings through symbolic queries\cite{Suyoto2008,Balke2016}; an approach to query a music
        video database with audio queries\cite{Gillet2007}.

    \subsection{Classification} 
        The \textit{classification} process consists in taking as input a
        music document and returning one or more labels. A popular multimodal
        classification task is the \textit{mood} or \textit{emotion}
        recognition\cite{Kim2010}, while an emerging one is \textit{genre}
        classification\cite{Li2004, Mayer2008, Mayer2008a, Mayer2010, Zhen2010, Mayer2011,Schindler2016, Oramas2018}.  Both these two tasks can take advantage of
        audio recordings, lyrics, cover arts and meta-tags.  Additionally,
        emotion recognition can exploit EEG data, while for genre
        classification one can use music video and generic text such as critic
        reviews. Usually, just one modality is considered in addition to audio
        recordings, but an interesting work\cite{Oramas2018} tries to exploit
        more than two modalities.  Other multimodal classification tasks found
        in the literature are:

        \begin{itemize}

            \item \textit{artist} identification, through lyrics and audio
                fusion\cite{Aryafar2014};

            \item \textit{derivative works} classification of youtube video
                through audio, video, titles and authors\cite{Smith2017};
                
            \item \textit{instrument} classification by exploiting audio recordings and
                performance video\cite{Slizovskaia2017, Lim2011};

            \item \textit{tonic} identification, that is: given an audio recording and
                the note level, find the tonic\cite{Sentuerk2013};

            \item \textit{expressive musical description}, which consists in
                associating a musical annotation to an audio recording by extracting
                features with the help of symbolic level\cite{Li2015}.

        \end{itemize}

    \subsection{Time-dependent representation} 
    	With \textit{time-dependent representation}, we mean the creation of a
        time-dependent description of the music data, created by merging and processing
        multiple modalities. Possibly the most studied task within this family is
        \textit{score-informed source separation}\cite{Mueller2015}, in which
        symbolic music data and audio recordings of a musical ensemble are used to
        create different audio recordings for each different instrument. A number
        of researchers have also tried to use audio and video recordings of a music
        performance or of a dancer to extract \textit{beat tracking}
        information\cite{Weinberg2009, Lim2010, Itohara2011,
        Berman2012, Ohkita2015}. An emerging task is \textit{piano tutoring}, which
        consists in the tracking of errors in a piano performance: to this end, 
        the audio recording, the knowledge about the instrument timbre and the symbolic
        score can be exploited\cite{Wang2017, Fukuda2015, Benetos2012, Mayor2009, Tsai2012,
        Abeser2013, Devaney2012}.
        Less studied tasks are:

        \begin{itemize}
        
        	\item \textit{music segmentation}, in which audio and video, lyrics or note
            	level can be exploited to identify the music piece structure \cite{Zhu2005,
                Cheng_2009, Gregorio2016};
        
            \item \textit{spatial transcription}, that is the inference, starting from audio and video,
            	of the note level of songs for fretted instruments, so that the resulting score includes the annotation of
                fingering\cite{Paleari2008, Hrybyk2010};

            \item \textit{onset} detection through audio and performer
                video\cite{Marenco2015} or rhythmic structure knowledge;

            \item \textit{chords} labeling, by comparing multiple audio recordings
                of the same work\cite{Konz2012};
                
            \item \textit{source association}, that is the detection of which
                player is active time by time by exploiting audio, video and music
                scores\cite{Li2017, Li2017a};

            \item \textit{multi-pitch} estimation, that is the transcription of parts being played simultaneously, with the help of
                performance video to detect play-nonplay activity of the
                various instruments\cite{Dinesh2017}.

        \end{itemize}

\section{Data pre-processing}\label{sec:preprocessing}
    \textit{Data pre-processing} is the elaboration of data to the end of
    transforming their representation to a more suitable format for the
    subsequent steps. We have identified a number of possible non-exclusive
    types of pre-processing :
    \begin{itemize}

        \item \textit{Synchronization}: the synchronization process described
            in \cref{sec:sync} is sometime used as pre-processing step to
            align multiple modalities; thus, the pre-processing itself can be
            multimodal. For example, in \textit{piano tutoring} and
            \textit{score-informed source separation}, an
            \textit{audio-to-score} alignment is performed; \textit{audio-to-audio} synchronization is a fundamental
            pre-processing step in tasks requiring comparison of multiple recordings of the same piece\cite{Konz2012}; \textit{audio-to-score} alignment is also used
            in several previously cited works\cite{Li2015, Balke2016, Suyoto2008, Gregorio2016};

        \item \textit{Feature extraction}: usually, music representations are
            not used as they are, but a number of features are extracted --~see
            \cref{sec:features}.

        \item Other pre-processing steps include:
            \begin{itemize}

                \item \textit{conversion} from one modality to the other,
                    such as in \textit{query-by-humming} --~which includes a
                    conversion from audio to the symbolic level~-- or in
                    \textit{audio-to-score} alignment where symbolic scores can be
                    converted to audio through a synthesis process.

                \item \textit{feature selection} through \textit{Linear
                    Discriminant Analysis} (LDA)\cite{Gillet2007} or
                    ReliefF\cite{Li2015}

                \item \textit{normalization} of the extracted
                    features\cite{Ohkita2015}

                \item \textit{source-separation} in lyrics-to-audio 
                	alignment and source association\cite{Li2017, Li2017a}

                \item chord labeling on audio only\cite{Konz2012}

                \item multi-pitch estimation on audio only\cite{Dinesh2017}

                \item video-based hand tracking\cite{Paleari2008}

                \item \textit{tf-idf}-based statistics --~see
                    \cref{sec:text-features}~-- adapted for
                    audio\cite{Aryafar2014}

            \end{itemize}

    \end{itemize}

    Finally, we think that a step worthy of a particular attention is the
    \textit{conversion to a common space} of the extracted features, to make
    them comparable. We will talk about this step in \cref{sec:conversion}.
    The accompanying online table (see \cref{foot:link}) contains a short description of the 
    pre-processing pipeline adopted in each cited paper.

\section{Feature extraction in multimodal approaches}\label{sec:features}

    Various types of features can be extracted from each modality. In this
    section, we provide a general description for audio, video, textual and
    symbolic score features. 
    
    \subsection{Audio features}
    
    This section is mainly written with reference to a previous review\cite{Alias2016}. Audio features can be broadly subdivided in \textit{physical} and \textit{perceptual}. 

    \subsubsection{Physical features} these can be computed in various domains,
    such as time, frequency or wavelet. Time-domain features can be
    computed directly on the digitally recorded audio signal and include
    \textit{zero-crossing rate}, \textit{amplitude}, \textit{rhythm} and
    \textit{power-based} features, such as the \textit{volume}, the
    \textit{MPEG-7 temporal centroid} or the \textit{beat histogram}.
    Frequency-domain features are the richest category; they are usually
    computed through a Short-Time Fourier Transform (STFT) or an autoregression
    analysis and can be subdivided in: \textit{autoregression-based},
    \textit{STFT-based} and \textit{brightness}, \textit{tonality},
    \textit{chroma} or \textit{spectrum shape} related. Features in the
    Wavelet-domain are computed after a Wavelet transform, which has the
    advantage of being able to represent discontinuous, finite, non-periodic or
    non-stationary functions. Image-domain features are computed through a
    graphic elaboration of the spectrogram, that is a matrix that can be
    represented as a one-channel image computed with the STFT; often,
    spectrogram is used as input for a convolutional neural network (CNN), which is trained to compute
    \textit{ad-hoc} features, which lack straightforward interpretation.
    
    \subsubsection{Perceptual features} these try to integrate human sound
    perception processing in the feature extraction stage or in the elaboration of
    physical audio features. Most of them aim at mapping certain measurements 
    to a perceptual-based scale anr/or metrics. For example, \textit{Mel Frequency
    Cepstral Coefficients} (MFCC) are derived by mapping the Fourier transform to 
    a Mel-scale, thus improving the coherence with human perception.
    \textit{Perceptual wavelet packets}\cite{Ntalampiras2009} employ a perceptually
    motivated critical-band based analysis to characterize each component of the
    spectrum using wavelet packets. \textit{Loudness} is computed from the Fourier transform with the aim of providing a psychophysically motivated measure of the
    intensity of a sound. 
    
    \subsection{Video and image features}

    This section is mainly written with reference to a previous work\cite{Ohkita2015}. 
    Video features used in the music domain are very similar to visual features
    used in general purpose video analysis. Image features can be based on the
    \textit{color space} (RGB or HSV), on \textit{edges} detection, on the
    \textit{texture} --~such as the LBP~--, or on the \textit{moment} of a
    region.  In video, motion detection is also possible and can be performed
    with \textit{background detection} and \textit{subtraction}, \textit{frame
    difference} and \textit{optical flow}. \textit{Object tracking} has been
    also used to detect hand movements, for example in \textit{piano-tutoring}
    applications.  \textit{Object tracking} can happen by exploiting the
    difference between frames of the detected object contours, by using
    deviations frame-to-frame of whole regions or generic features. In video, one can also detect \textit{shots}, for example by
    analyzing the variation of the color histograms in the video frames,
    using the Kullbach-Leibler distance\cite{Brezeale2008} or other metrics.

    In genre and mood related analysis, other features can also be
    exploited\cite{Wang2015}. The use of \textit{tempo} is essential to
    express emotions in video clips, and can be analyzed through features related to motion and length of
    video shots. Another relevant factor is lighting,
    that can be measured through brightness-based features. Colors have an affective meaning too, and color features are consequently
    useful for genre or emotion recognition.

    Finally, images can also be used as they are as input of CNNs.

    \subsection{Text features}\label{sec:text-features}

    This section is written with reference to a previous review\cite{Croft2015}. 
    The most common text representations are based on \textit{tf-idf}. In this
    context, tf$(d, t)$ is the \textit{term frequency} and is computed as the
    number of occurrences of a term $t$ in a document $d$. Instead, idf$(d, t)$
    is a short for \textit{inverse document frequency} and is needed to
    integrate the discrimination power of the term $t$ for the document $d$,
    considering the whole collection; it is related to the inverse ratio between the
    number of documents containing $t$ at least once and the total number of
    documents in the considered collection:
    \begin{equation} 
    \text{idf} = \frac{\text{docs~in~collection}}{\text{docs~containing~t}}
    \end{equation}
    Usually, \textit{tf-idf} takes the following form:
    \begin{equation} 
    \text{tf-idf}(d, t) = \text{tf}(d, t) \times \log[\text{idf}(d, t)]
    \end{equation}
    
    Features based on \textit{tf-idf} are often used in Bag-of-Words (BoW) models, where
    each document is represented as a list of words, without taking care of the
    cardinality and order of words. In order to make BoW and \textit{tf-idf}
    models effective, a few preliminary steps are usually performed, such as
    \textit{punctuation} and \textit{stop-words} removal and \textit{stemming}.
    More sophisticated methods are also available, allowing for topic- or
    semantics-based analysis, such as \textit{Latent Dirichlet Allocation}
    (LDA), \textit{Latent Semantic Analysis} (LSA), \textit{Explicit Semantic
    Analysis} (ESA)\cite{Gabrilovich2007} and CNN feature extraction.

    For lyrics analysis, other types of features can be extracted, like rhymes
    or positional features. Finally, when the available text is limited, one can
    extend it with a semantic approach consisting in \textit{knowledge
    boosting}\cite{Oramas2018}.

    \subsection{Symbolic score features}

    Symbolic music scores have been rarely used in feature extraction
    approaches. Most of the papers which deal with symbolic scores use
    MIDI-derived representations, such as the pianoroll\cite{Mueller2015} or
    inter-onset intervals (IOI)\cite{Gregorio2016}. To the end of audio-symbolic
    comparison, one can compute chromograms, that are also computable from
    the audio modality alone. However a number of representation exist and have been
    tested in Music Information Retrieval applications, such as \textit{pitch
    histograms}, \textit{Generalized Pitch Interval Representation (GPIR)},
    \textit{Spiral Array}, \textit{Rizo-Iñesta trees}, \textit{Pinto graphs},
    \textit{Orio-Rodà graphs} and others. A brief review of the
    music symbolic level representations is provided in a previous work\cite{Simonetta2018}.
    
\section{Conversion to common space}\label{sec:conversion}
    The conversion of the extracted features to a \textit{common space}  
    is often a mandatory step in \textit{early fusion} approaches. Nevertheless,
    almost no authors emphasize this aspect. Thus, we think that greater attention
    should be posed on this step of the pre-processing pipeline.
    
    The conversion to a common space consists in the mapping of the
    features coming from different modalities to a new space where they are
    comparable. This can be needed in single-modality approaches too, when
    the features refer to very different characteristics of the signal. Indeed,
    many papers describe techniques which include a mapping of the features
    to a common space, both in the pre-processing and in the processing
    stages, but no particular attention to the conversion itself. Common
    methods include:

    \begin{itemize}

        \item \textit{normalization}, that is the most basic approach;

        \item \textit{conversion from one modality to another}, so that
            features can be computed in the same units;

        \item \textit{machine learning algorithms} such as CNNs or SVMs:
            SVMs compute the best parameters for a kernel function that is
            used to transform the data into a space where they are more
            easily separable; CNNs, instead, can be trained to represent
            each input modality in a space such that the last network
            layers can use as input the concatenation of these
            representations;

        \item \textit{dimensionality reduction} algorithms, which usually
            search for a new space where data samples are representable
            with a fewer number of dimensions without losing the ability
            to separate them; examples are \textit{Principal Component
            Analysis} (PCA) and \textit{Linear Discriminant Analysis} (LDA).

    \end{itemize}

    It must be said that some types of features are suitable for multimodal
    fusion without any conversion step. For example, \textit{chroma
    features} can be computed from both the audio recordings and the
    symbolic scores and thus can be compared with no additional processing.
    
    A possible categorization of the conversion to common space methods is between
    \textit{coordinated} and \textit{joint}: in the former type, the mapping function
    takes as input a unimodal representation, while in the latter type it takes
    as input a multimodal representation\cite{Baltrusaitis2018}. In other words,
    \textit{coordinated} conversion learns to map each modality to a new space trying to
    minimize the distance between the various descriptions of the same object, while
    \textit{joint} conversion learns the best mapping function which uses all the modalities
    and optimizes the subsequent steps --~e.g. SVM.

\section{Information fusion approaches}\label{sec:fusion}
    Two major information fusion approaches exist: \textit{early fusion} and
    \textit{late fusion} --~see \cref{fig:fusion}. Some authors also report a
    \textit{hybrid} approach\cite{Atrey2010}, which consists in fusing
    information both in a \textit{early} and \textit{late} fashion and in
    adding a further step to fuse the output of the two approaches.
    Nevertheless, we did not find any existing application to the music domain.
    Before discussing in detail the two approaches, we recall that no fusion is usually needed in \textit{similarity} tasks, but
    just a comparison of the various modalities and, thus, a conversion to a
    common space. The accompanying online table (see \cref{foot:link}) contains a short 
    description of the fusion approach used in all the cited papers. To our
    understanding the main difference between \textit{early} and \textit{late}
    fusion is about their efficiency and ease of development; however authors
    disagree about which one is the more effective.

    \subsection{Early fusion}
        \textit{Early fusion} consists in the fusion of the
        features of all the modalities, using them as input in one single processing
        algorithm. Although the development of such techniques is more
        straightforward, they need a more careful treatment because the features
        extracted from various modalities are not always directly comparable.

        To the end of \textit{synchronization}, the most used approach exploits
        Dynamic Time Warping algorithms (DTW)\cite{Mueller2007}. 
        DTW is a
        well-known technique based on a similarity matrix between two sorted
        sets of points, for example two time-sequences. By using a dynamic
        programming algorithm, one can exploit the similarity matrix to find
        the best path which connects the first point in one modality to the
        last point in the same modality and which satisfies certain conditions.
        This path will indicate the corresponding points between the two
        modalities. 
        Other common methods for synchronization purposes are
        Hidden Markov Models (HMMs)\cite{Fujihara2012, Gregorio2016} where
        hidden states represent points in one modality and observations represent points
        in a second modality; this is particularly effective for real-time
        alignment or generic sequence fusion such as in \textit{time-dependent
        descriptions}.

        Aside HMMs, many additional machine learning\cite{Bishop2007} approaches are
        used to perform early fusion: Support Vector Machines (SVMs),
        Gaussian Mixture Models (GMMs), Convolutional Neural Networks (CNNs)
        and Particle Filters are the most used techniques. 

\begin{figure}[]
        \centerline{\includegraphics[width=0.5\textwidth]{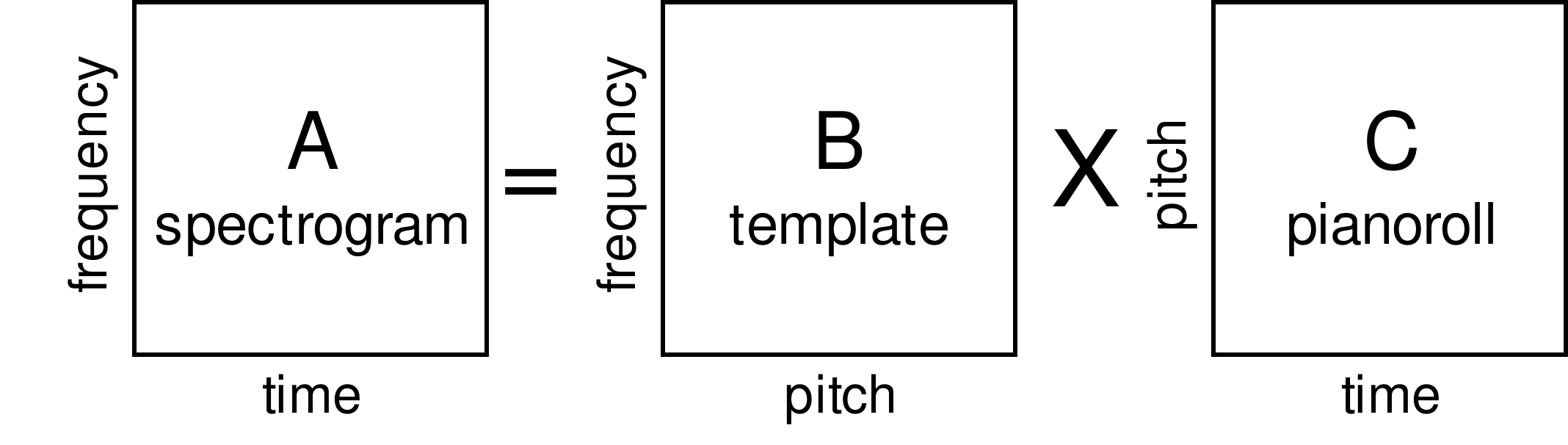}}
        \caption{Exemplification of Non-negative Matrix Factorization for music transcription.}
        \label{fig:nmf}
    \end{figure}

        Another interesting method is Non-negative Matrix Factorization (NMF),
        through which audio and symbolic scores can be exploited to the end of
        precise performance transcription, as in \textit{score-informed source
        separation} and \textit{piano tutoring} applications\cite{Mueller2015}.
        In NMF, a matrix $A$ is decomposed in two components $C$ and $B$, so
        that $A = B \times C$. If $A$ is a spectrogram 
        and $B$ is a \textit{template matrix} dependent on the
        instrumentation, then we can think to $C$ as a pianoroll matrix --~see \cref{fig:nmf}.  Consequently, one can use an optimization algorithm
        to minimize the function $f(B, C) = A - B \times C$, by initializing
        $C$ with a symbolic score; at the end of the optimization, $C$ will be
        a precise transcription of the performance contained in $A$.
        
        Finally, feature fusion can also happen at the feature selection
        stage\cite{Li2015, Aryafar2014}.

    \subsection{Late fusion}
        Unlike \textit{early fusion}, \textit{late fusion} is the fusion of the
        output of various
        \textit{ad-hoc} algorithms, one for each modality. It is also called
        \textit{decision-level} fusion, even if a decision process is not
        mandatory. The main advantage of \textit{late fusion} is that it allows
        for a more adjustable processing of each modality. However, it is
        usually more demanding in terms of development costs.

        In \textit{classification} and \textit{time-dependent description}
        tasks, the most used types of late fusion are \textit{rule-based}.
        Rules can include voting procedures\cite{Konz2012, Li2017a}, linear and
        weighted combinations\cite{Lim2011, Degara2010}, maximum and minimum
        operations\cite{Lim2011, Degara2010}.  Many authors have developed sophisticated algorithms to execute this step, such as in \textit{beat tracking},
        \textit{piano tutoring} and \textit{structural segmentation}\cite{Cheng_2009}, multi-pitch estimation\cite{Dinesh2017} and tonic identification\cite{Sentuerk2013}.

        In \textit{synchronization} tasks, instead, no \textit{late-fusion}
        approach is possible, since the task consists in creating
        associations between points in different modalities and, thus, the
        process must take as input all the modalities, eventually in some
        common representation.
        
\section{Future directions}\label{sec:future}

    In this paper, we have analyzed the literature on multimodal music
    information processing and retrieval. Based on our study, we propose
    the following concluding remarks.
    
    First of all, we note the unavailability of datasets of suitable size. This issue is usually addressed with various methods such as \textit{co-learning}
    approaches\cite{Baltrusaitis2018}, which has the side-effect of impoverishing the
    goodness of the developed algorithms. Although a few datasets have been 
    recently created\cite{MeseguerBrocal2018, Maestre2017, Li2018, Oramas2018}, a great effort should still be carried out in this direction. Indeed, existing multimodal music datasets are usually characterized by limited sizes and only rarely include a wide range of modalities. However an exhaustive list of the available datasets is out of the scope of this paper. 
    We argue that this limit is due to two main reasons: first, the precise alignment of various
    modalities is a hard computational task and should be controlled by
    human supervision; second, no largely adopted standard exists for
    multimodal music representation. About the first point, more
    effort should be devoted to the development of algorithms for the
    alignment of various sequences. The representation of the intrinsic
    music multimodality, instead, is faced by the \textit{IEEE 1599}\footnote{IEEE 1599 website: \url{http://ieee1599.lim.di.unimi.it/}}
    standard and the \textit{Music Encoding Initiative}\footnote{MEI website: \url{https://music-encoding.org/}}; moreover, the \textit{W3C}
    group is currently working on a new standard with the purpose of
    enriching \textit{MusicXML} with multimodal information\footnote{W3C music notation group website: \url{https://www.w3.org/community/music-notation/}}. 
    The course of history described
    in\cref{sec:intro} and the rapid technology advancements of our times suggest that new representation modalities
    could be needed in the future and that multimodal representation
    standards should also focus on this challenge.
    
    Another challenge that multimodal music researchers should face in the next years is
    the exploration of various techniques already used in multimodal
    processing of multimedia data, that have not been tested in the musical domain.
    According to previous surveys\cite{Atrey2010, Baltrusaitis2018}, multimodal methods
    never applied to the music domain include: the hybrid approach, the
    Dempster-Shafer theory, Kalman filters, the maximum entropy model, Multiple Kernel
    Learning and Graphical Models. Moreover, we have 
    found only one paper in which the information fusion happens during
    the feature extraction itself\cite{Gregorio2016} and not afterwards.
    This approach should be explored more deeply.

    Finally, we suggest that the conversion to common space should be more 
    rigorously addressed.
    To this end, transfer learning technologies could be explored
    towards forming a synergistic feature space able to meaningfully
    represent multiple modalities\cite{Pan2010,Ntalampiras2017}. Such a
    direction may include the use of an existing feature space
    characterizing a specific modality, or the creation of a new one where
    multiple modalities are represented. Such a space could satisfy several
    desired properties, such as sparseness, reduced dimensionality, and so on.

\balance
\bibliographystyle{IEEEtran}
\bibliography{bibliography}

\begin{thebibliography}{10}
\providecommand{\url}[1]{#1}
\csname url@samestyle\endcsname
\providecommand{\newblock}{\relax}
\providecommand{\bibinfo}[2]{#2}
\providecommand{\BIBentrySTDinterwordspacing}{\spaceskip=0pt\relax}
\providecommand{\BIBentryALTinterwordstretchfactor}{4}
\providecommand{\BIBentryALTinterwordspacing}{\spaceskip=\fontdimen2\font plus
\BIBentryALTinterwordstretchfactor\fontdimen3\font minus
  \fontdimen4\font\relax}
\providecommand{\BIBforeignlanguage}[2]{{%
\expandafter\ifx\csname l@#1\endcsname\relax
\typeout{** WARNING: IEEEtran.bst: No hyphenation pattern has been}%
\typeout{** loaded for the language `#1'. Using the pattern for}%
\typeout{** the default language instead.}%
\else
\language=\csname l@#1\endcsname
\fi
#2}}
\providecommand{\BIBdecl}{\relax}
\BIBdecl

\bibitem{Bent2001}
I.~D. Bent, D.~W. Hughes, R.~C. Provine, R.~Rastall, A.~Kilmer, D.~Hiley,
  J.~Szendrei, T.~B. Payne, M.~Bent, and G.~Chew, ``Notation,'' in \emph{Grove
  Music Online}.\hskip 1em plus 0.5em minus 0.4em\relax Oxford University
  Press, 2001.

\bibitem{Brown2001}
H.~M. Brown, E.~Rosand, R.~Strohm, M.~Noiray, R.~Parker, A.~Whittall,
  R.~Savage, and B.~Millington, ``Opera (i),'' in \emph{Grove Music
  Online}.\hskip 1em plus 0.5em minus 0.4em\relax Oxford University Press,
  2001.

\bibitem{Mumma2003}
G.~Mumma, H.~Rye, B.~Kernfeld, and C.~Sheridan, ``Recording,'' in \emph{Grove
  Music Online}.\hskip 1em plus 0.5em minus 0.4em\relax Oxford University
  Press, 2003.

\bibitem{Cook2018}
D.~A. Cook and R.~Sklar, ``History of the motion picture,'' in \emph{Britannica
  Academic}.\hskip 1em plus 0.5em minus 0.4em\relax Encyclopædia Britannica,
  2018.

\bibitem{Kitahara2010}
T.~Kitahara, ``Mid-level representations of musical audio signals for music
  information retrieval,'' in \emph{Advances in Music Information
  Retrieval}.\hskip 1em plus 0.5em minus 0.4em\relax Springer Berlin
  Heidelberg, 2010, pp. 65--91.

\bibitem{Vinet2004}
H.~Vinet, ``{The Representation Levels of Music Information},'' in
  \emph{Computer Music Modeling and Retrieval}, U.~K. Wiil, Ed.\hskip 1em plus
  0.5em minus 0.4em\relax Berlin, Heidelberg: Springer Berlin Heidelberg, 2004,
  pp. 193--209.

\bibitem{Pachet2005}
F.~Pachet, ``Musical metadata and knowledge management,'' in \emph{Encyclopedia
  of Knowledge Management, Second Edition}, D.~G. Schwartz and D.~Te'eni,
  Eds.\hskip 1em plus 0.5em minus 0.4em\relax {IGI} Global, 2005, pp.
  1192--1199.

\bibitem{Deutsch2013a}
D.~Deutsch, \emph{The Psychology of Music (third Edition)}, third edition~ed.,
  D.~Deutsch, Ed.\hskip 1em plus 0.5em minus 0.4em\relax Academic Press, 2013.

\bibitem{Atrey2010}
P.~K. Atrey, M.~A. Hossain, A.~E. Saddik, and M.~S. Kankanhalli, ``Multimodal
  fusion for multimedia analysis: A survey,'' \emph{Multimedia Systems},
  vol.~16, no.~6, pp. 345--379, Apr. 2010.

\bibitem{Baltrusaitis2018}
T.~Baltrusaitis, C.~Ahuja, and L.-P. Morency, ``Multimodal machine learning: A
  survey and taxonomy,'' \emph{{IEEE} Transactions on Pattern Analysis and
  Machine Intelligence}, pp. 1--1, 2018.

\bibitem{Minsky1991}
M.~Minsky, ``Logical versus analogical or symbolic versus connectionist or neat
  versus scruffy,'' \emph{{AI} Magazine}, vol.~12, no.~2, pp. 34--51, 1991.

\bibitem{Essid2012}
S.~Essid and G.~Richard, ``\BIBforeignlanguage{eng}{Fusion of multimodal
  information in music content analysis},'' in
  \emph{\BIBforeignlanguage{eng}{Multimodal Music Processing}}.\hskip 1em plus
  0.5em minus 0.4em\relax Schloss Dagstuhl - Leibniz-Zentrum fuer Informatik
  GmbH, Wadern/Saarbruecken, Germany, 2012.

\bibitem{Schedl2014}
M.~Schedl, E.~G\'{o}mez, and J.~Urbano, ``Music information retrieval: Recent
  developments and applications,'' \emph{Foundations and Trends in Information
  Retrieval}, vol.~8, no. 2-3, pp. 127--261, 2014.

\bibitem{Tzanetakis2002}
G.~Tzanetakis and P.~Cook, ``Musical genre classification of audio signals,''
  \emph{IEEE Transactions on Speech and Audio Processing}, vol.~10, no.~5, pp.
  293--302, July 2002.

\bibitem{Nanni2016}
L.~Nanni, Y.~M.~G. Costa, A.~Lumini, M.~Y. Kim, and S.~Baek, ``Combining visual
  and acoustic features for music genre classification,'' \emph{Expert Syst.
  Appl.}, vol.~45, pp. 108--117, 2016.

\bibitem{Zhonghua2013}
L.~Zhonghua, ``Multimodal music information retrieval: From content analysis to
  multimodal fusion,'' Ph.D. dissertation, 2013.

\bibitem{Mueller2015}
M.~M{ü}ller, \emph{Fundamentals of Music Processing: Audio, Analysis,
  Algorithms, Applications}, 1st~ed.\hskip 1em plus 0.5em minus 0.4em\relax
  Springer Publishing Company, Incorporated, 2015.

\bibitem{Fujihara2012}
H.~Fujihara and M.~Goto, ``{Lyrics-to-Audio Alignment and its Application},''
  in \emph{Multimodal Music Processing}, ser. Dagstuhl Follow-Ups,
  M.~M{\"u}ller, M.~Goto, and M.~Schedl, Eds.\hskip 1em plus 0.5em minus
  0.4em\relax Dagstuhl, Germany: Schloss Dagstuhl--Leibniz-Zentrum fuer
  Informatik, 2012, vol.~3, pp. 23--36.

\bibitem{Dorfer2017}
M.~Dorfer, A.~Arzt, and G.~Widmer, ``Learning audio-sheet music correspondences
  for score identification and offline alignment,'' in \emph{Proceedings of the
  18th International Society for Music Information Retrieval Conference,
  {ISMIR} 2017, Suzhou, China}, S.~J. Cunningham, Z.~Duan, X.~Hu, and
  D.~Turnbull, Eds., 2017, pp. 115--122.

\bibitem{Kotsifakos2012}
A.~Kotsifakos, P.~Papapetrou, J.~Hollmén, D.~Gunopulos, and V.~Athitsos, ``A
  survey of query-by-humming similarity methods,'' in \emph{Proceedings of the
  5th International Conference on PErvasive Technologies Related to Assistive
  Environments}, ser. PETRA '12.\hskip 1em plus 0.5em minus 0.4em\relax New
  York, NY, USA: ACM, 2012.

\bibitem{MIREX2016}
\BIBentryALTinterwordspacing
{MIREX Community}. (2016) 2016:query by singing/humming. [Online]. Available:
  \url{https://www.music-ir.org/mirex/wiki/2016:Query\_by\_Singing/Humming}
\BIBentrySTDinterwordspacing

\bibitem{Bonnin2014}
G.~Bonnin and D.~Jannach, ``Automated generation of music playlists: Survey and
  experiments,'' \emph{ACM Comput. Surv.}, vol.~47, no.~2, pp. 26:1--26:35,
  Nov. 2014.

\bibitem{Deshmukh2018}
P.~Deshmukh and G.~Kale, ``{A Survey of Music Recommendation System},''
  \emph{International Journal of Scientific Research in Computer Science,
  Engineering and Information Technology (IJSRCSEIT)}, vol.~3, no.~3, pp.
  1721--1729, Mar. 2018.

\bibitem{Mueller2019}
M.~M{ü}ller, A.~Arzt, S.~Balke, M.~Dorfer, and G.~Widmer, ``Cross-modal music
  retrieval and applications: An overview of key methodologies,'' \emph{IEEE
  Signal Processing Magazine}, vol.~36, no.~1, pp. 52--62, Jan. 2019.

\bibitem{Correya2018}
A.~Correya, R.~Hennequin, and M.~Arcos, ``{Large-Scale Cover Song Detection in
  Digital Music Libraries Using Metadata, Lyrics and Audio Features},''
  \emph{Arxiv E-prints}, Aug. 2018.

\bibitem{Suyoto2008}
I.~S.~H. Suyoto, A.~L. Uitdenbogerd, and F.~Scholer, ``Searching musical audio
  using symbolic queries,'' \emph{{IEEE} Transactions on Audio, Speech, and
  Language Processing}, vol.~16, no.~2, pp. 372--381, Feb. 2008.

\bibitem{Balke2016}
S.~Balke, V.~Arifi-M{ü}ller, L.~Lamprecht, and M.~M{ü}ller, ``Retrieving
  audio recordings using musical themes,'' in \emph{2016 IEEE International
  Conference on Acoustics, Speech and Signal Processing (ICASSP)}, Mar. 2016,
  pp. 281--285.

\bibitem{Gillet2007}
O.~Gillet, S.~Essid, and G.~Richard, ``On the correlation of automatic audio
  and visual segmentations of music videos,'' \emph{Ieee Transactions on
  Circuits and Systems for Video Technology}, vol.~17, no.~3, pp. 347--355,
  2007-03.

\bibitem{Kim2010}
Y.~E. Kim, E.~M. Schmidt, R.~Migneco, B.~G. Morton, P.~Richardson, J.~J. Scott,
  J.~A. Speck, and D.~Turnbull, ``State of the art report: Music emotion
  recognition: {A} state of the art review,'' in \emph{Proceedings of the 11th
  International Society for Music Information Retrieval Conference, {ISMIR}
  2010, Utrecht, Netherlands, August 9-13, 2010}, J.~S. Downie and R.~C.
  Veltkamp, Eds.\hskip 1em plus 0.5em minus 0.4em\relax International Society
  for Music Information Retrieval, 2010, pp. 255--266.

\bibitem{Li2004}
T.~Li and M.~Ogihara, ``Music artist style identification by semi-supervised
  learning from both lyrics and content,'' in \emph{Proceedings of the 12th
  Annual ACM International Conference on Multimedia}, ser. MULTIMEDIA
  '04.\hskip 1em plus 0.5em minus 0.4em\relax New York, NY, USA: ACM, 2004, pp.
  364--367.

\bibitem{Mayer2008}
R.~Mayer, R.~Neumayer, and A.~Rauber, ``Combination of audio and lyrics
  features for genre classification in digital audio collections,'' in
  \emph{Proceedings of the 16th ACM International Conference on Multimedia},
  ser. MM '08.\hskip 1em plus 0.5em minus 0.4em\relax New York, NY, USA: ACM,
  2008, pp. 159--168.

\bibitem{Mayer2008a}
------, ``Rhyme and style features for musical genre classification by song
  lyrics,'' in \emph{{ISMIR} 2008, 9th International Conference on Music
  Information Retrieval, Drexel University, Philadelphia, PA, USA, September
  14-18, 2008}, 2008, pp. 337--342.

\bibitem{Mayer2010}
R.~Mayer and A.~Rauber, \emph{Multimodal Aspects of Music Retrieval: Audio,
  Song Lyrics -- and Beyond?}\hskip 1em plus 0.5em minus 0.4em\relax Berlin,
  Heidelberg: Springer Berlin Heidelberg, 2010, pp. 333--363.

\bibitem{Zhen2010}
C.~Zhen and J.~Xu, ``Multi-modal music genre classification approach,'' in
  \emph{Proc. 3rd Int. Conf. Computer Science and Information Technology},
  vol.~8, Jul. 2010, pp. 398--402.

\bibitem{Mayer2011}
R.~Mayer and A.~Rauber, ``Musical genre classification by ensembles of audio
  and lyrics features,'' in \emph{Proceedings of the 12th International Society
  for Music Information Retrieval Conference (ISMIR 2011)}.\hskip 1em plus
  0.5em minus 0.4em\relax University of Miami, 2011, pp. 675--680, vortrag:
  12th International Society for Music Information Retrieval Conference (ISMIR
  2011), Miami; 2011-10-24 -- 2011-10-28.

\bibitem{Schindler2016}
A.~Schindler and A.~Rauber, ``Harnessing music-related visual stereotypes for
  music information retrieval,'' \emph{{ACM} Transactions on Intelligent
  Systems and Technology}, vol.~8, no.~2, pp. 1--21, Oct. 2016.

\bibitem{Oramas2018}
S.~Oramas, F.~Barbieri, O.~Nieto, and X.~Serra, ``Multimodal deep learning for
  music genre classification,'' \emph{Transactions of the International Society
  for Music Information Retrieval}, vol.~1, no.~1, pp. 4--21, 2018.

\bibitem{Aryafar2014}
K.~Aryafar and A.~Shokoufandeh, ``Multimodal music and lyrics fusion classifier
  for artist identification,'' in \emph{2014 13th International Conference on
  Machine Learning and Applications}.\hskip 1em plus 0.5em minus 0.4em\relax
  {IEEE}, Dec. 2014.

\bibitem{Smith2017}
J.~B.~L. Smith, M.~Hamasaki, and M.~Goto, ``Classifying derivative works with
  search, text, audio and video features,'' in \emph{Proc. IEEE Int. Conf.
  Multimedia and Expo (ICME)}, Jul. 2017, pp. 1422--1427.

\bibitem{Slizovskaia2017}
O.~Slizovskaia, E.~Gómez, and G.~Haro, ``Musical instrument recognition in
  user-generated videos using a multimodal convolutional neural network
  architecture,'' in \emph{Proceedings of the 2017 {ACM} on International
  Conference on Multimedia Retrieval - {ICMR} {\textquotesingle}17}.\hskip 1em
  plus 0.5em minus 0.4em\relax {ACM} Press, 2017.

\bibitem{Lim2011}
A.~Lim, K.~Nakamura, K.~Nakadai, T.~Ogata, and H.~G. Okuno, ``Audio-visual
  musical instrument recognition,'' 2011.

\bibitem{Sentuerk2013}
S.~Sentürk, S.~Gulati, and X.~Serra, ``Score informed tonic identification for
  makam music of turkey,'' in \emph{Proceedings of the 14th International
  Society for Music Information Retrieval Conference, {ISMIR} 2013, Curitiba,
  Brazil, November 4-8, 2013}, A.~de~Souza Britto~Jr., F.~Gouyon, and S.~Dixon,
  Eds., 2013, pp. 175--180.

\bibitem{Li2015}
P.-C. Li, L.~Su, Y.-H. Yang, and A.~W.~Y. Su, ``Analysis of expressive musical
  terms in violin using score-informed and expression-based audio features,''
  in \emph{Proceedings of the 16th International Society for Music Information
  Retrieval Conference, {ISMIR} 2015, M{\'{a}}laga, Spain, October 26-30,
  2015}, M.~M{\"{u}}ller and F.~Wiering, Eds., 2015, pp. 809--815.

\bibitem{Weinberg2009}
G.~Weinberg, A.~Raman, and T.~Mallikarjuna, ``Interactive jamming with shimon:
  A social robotic musician,'' in \emph{Proceedings of the 4th ACM/IEEE
  International Conference on Human Robot Interaction}, ser. HRI '09.\hskip 1em
  plus 0.5em minus 0.4em\relax New York, NY, USA: ACM, 2009, pp. 233--234.

\bibitem{Lim2010}
A.~Lim, T.~Mizumoto, L.~Cahier, T.~Otsuka, T.~Takahashi, K.~Komatani, T.~Ogata,
  and H.~G. Okuno, ``Robot musical accompaniment: integrating audio and visual
  cues for real-time synchronization with a human flutist,'' in \emph{Proc.
  IEEE/RSJ Int. Conf. Intelligent Robots and Systems}, Oct. 2010, pp.
  1964--1969.

\bibitem{Itohara2011}
T.~Itohara, T.~Otsuka, T.~Mizumoto, T.~Ogata, and H.~G. Okuno,
  ``Particle-filter based audio-visual beat-tracking for music robot ensemble
  with human guitarist,'' in \emph{Proc. IEEE/RSJ Int. Conf. Intelligent Robots
  and Systems}, Sep. 2011, pp. 118--124.

\bibitem{Berman2012}
D.~R. Berman, ``{AVISARME: Audio Visual Synchronization Algorithm for a Robotic
  Musician Ensemble},'' Ph.D. dissertation, University of Maryland, 2012.

\bibitem{Ohkita2015}
M.~Ohkita, Y.~Bando, Y.~Ikemiya, K.~Itoyama, and K.~Yoshii, ``Audio-visual beat
  tracking based on a state-space model for a music robot dancing with
  humans,'' in \emph{Proc. IEEE/RSJ Int. Conf. Intelligent Robots and Systems
  (IROS)}, Sep. 2015, pp. 5555--5560.

\bibitem{Wang2017}
S.~Wang, S.~Ewert, and S.~Dixon, ``Identifying missing and extra notes in piano
  recordings using score-informed dictionary learning,'' \emph{{IEEE}/{ACM}
  Transactions on Audio, Speech, and Language Processing}, vol.~25, no.~10, pp.
  1877--1889, 2017.

\bibitem{Fukuda2015}
T.~Fukuda, Y.~Ikemiya, K.~Itoyama, and K.~Yoshii, ``A score-informed piano
  tutoring system with mistake detection and score simplification,'' in
  \emph{Proc of the Sound and Music Computing Conference (SMC)}.\hskip 1em plus
  0.5em minus 0.4em\relax Zenodo, 2015.

\bibitem{Benetos2012}
E.~Benetos, A.~Klapuri, and S.~Dixon, ``{Score-informed transcription for
  automatic piano tutoring},'' in \emph{European Signal Processing Conference},
  2012, pp. 2153--2157.

\bibitem{Mayor2009}
O.~Mayor, J.~Bonada, and A.~Loscos, ``Performance analysis and scoring of the
  singing voice,'' in \emph{AES 35th International Conference: Audio for
  Games}, 2009.

\bibitem{Tsai2012}
W.~Tsai and H.~Lee, ``Automatic evaluation of karaoke singing based on pitch,
  volume, and rhythm features,'' \emph{and Language Processing IEEE
  Transactions on Audio, Speech}, vol.~20, no.~4, pp. 1233--1243, May 2012.

\bibitem{Abeser2013}
J.~Abe{\ss}er, J.~Hasselhorn, C.~Dittmar, A.~Lehmann, and S.~Grollmisch,
  ``Automatic quality assessment of vocal and instrumental performances of
  ninth-grade and tenth-grade pupils,'' in \emph{International Symposium on
  Computer Music Multidisciplinary Research}, 2013.

\bibitem{Devaney2012}
J.~Devaney, M.~I. Mandel, and I.~Fujinaga, ``A study of intonation in
  three-part singing using the automatic music performance analysis and
  comparison toolkit {(AMPACT)},'' in \emph{Proceedings of the 13th
  International Society for Music Information Retrieval Conference, {ISMIR}
  2012, Mosteiro S.Bento Da Vit{\'{o}}ria, Porto, Portugal, October 8-12,
  2012}, F.~Gouyon, P.~Herrera, L.~G. Martins, and M.~M{\"{u}}ller, Eds.\hskip
  1em plus 0.5em minus 0.4em\relax {FEUP} Edi{\c{c}}{\~{o}}es, 2012, pp.
  511--516.

\bibitem{Zhu2005}
Y.~Zhu, K.~Chen, and Q.~Sun, ``Multimodal content-based structure analysis of
  karaoke music,'' in \emph{Proceedings of the 13th Annual ACM International
  Conference on Multimedia}, ser. MULTIMEDIA '05.\hskip 1em plus 0.5em minus
  0.4em\relax New York, NY, USA: ACM, 2005, pp. 638--647.

\bibitem{Cheng_2009}
H.-T. Cheng, Y.-H. Yang, Y.-C. Lin, and H.~H. Chen, ``Multimodal structure
  segmentation and analysis of music using audio and textual information,'' in
  \emph{2009 {IEEE} International Symposium on Circuits and Systems}.\hskip 1em
  plus 0.5em minus 0.4em\relax {IEEE}, May 2009.

\bibitem{Gregorio2016}
J.~Gregorio and Y.~Kim, ``Phrase-level audio segmentation of jazz
  improvisations informed by symbolic data.'' in \emph{Proceedings of the 17th
  International Society for Music Information Retrieval Conference, {ISMIR}
  2016, New York City, United States}, 2016, pp. 482--487.

\bibitem{Paleari2008}
M.~Paleari, B.~Huet, A.~Schutz, and D.~Slock, ``A multimodal approach to music
  transcription,'' in \emph{Proc. 15th IEEE Int. Conf. Image Processing}, Oct.
  2008, pp. 93--96.

\bibitem{Hrybyk2010}
A.~Hrybyk, ``Combined audio and video analysis for guitar chord
  identification,'' in \emph{11th International Society for Music Information
  Retrieval Conference (ISMIR 2010)}, 2010.

\bibitem{Marenco2015}
B.~Marenco, M.~Fuentes, F.~Lanzaro, M.~Rocamora, and A.~Gómez, ``A multimodal
  approach for percussion music transcription from audio and video,''
  \emph{Progress in Pattern Recognition, Image Analysis, Computer Vision, and
  Applications}, Jan. 2015.

\bibitem{Konz2012}
V.~Konz and M.~M{ü}ller, ``A cross-version approach for harmonic analysis of
  music recordings,'' in \emph{Multimodal Music Processing}, ser. Dagstuhl
  Follow-Ups, M.~M{\"{u}}ller, M.~Goto, and M.~Schedl, Eds.\hskip 1em plus
  0.5em minus 0.4em\relax Schloss Dagstuhl - Leibniz-Zentrum fuer Informatik,
  Germany, 2012, vol.~3, pp. 53--72.

\bibitem{Li2017}
B.~Li, C.~Xu, and Z.~Duan, ``Audiovisual source association for string
  ensembles through multi-modal vibrato analysis,'' \emph{Proc. Sound and Music
  Computing (smc)}, 2017.

\bibitem{Li2017a}
B.~Li, K.~Dinesh, Z.~Duan, and G.~Sharma, ``See and listen: Score-informed
  association of sound tracks to players in chamber music performance videos,''
  in \emph{Proc. Speech and Signal Processing (ICASSP) 2017 IEEE Int. Conf.
  Acoustics}, Mar. 2017, pp. 2906--2910.

\bibitem{Dinesh2017}
K.~Dinesh, B.~Li, X.~Liu, Z.~Duan, and G.~Sharma, ``Visually informed
  multi-pitch analysis of string ensembles,'' in \emph{Proc. Speech and Signal
  Processing (ICASSP) 2017 IEEE Int. Conf. Acoustics}, Mar. 2017, pp.
  3021--3025.

\bibitem{Alias2016}
F.~Alías, J.~Socoró, and X.~Sevillano, ``A review of physical and perceptual
  feature extraction techniques for speech, music and environmental sounds,''
  \emph{Applied Sciences}, vol.~6, no.~5, p. 143, May 2016.

\bibitem{Ntalampiras2009}
S.~Ntalampiras, I.~Potamitis, and N.~Fakotakis, ``Exploiting temporal feature
  integration for generalized sound recognition,'' \emph{EURASIP Journal on
  Advances in Signal Processing}, vol. 2009, no.~1, p. 807162, Dec 2009.

\bibitem{Brezeale2008}
D.~Brezeale and D.~J. Cook, ``Automatic video classification: A survey of the
  literature,'' \emph{IEEE Transactions on Systems, Man, and Cybernetics, Part
  C (Applications and Reviews)}, vol.~38, no.~3, pp. 416--430, May 2008.

\bibitem{Wang2015}
S.~Wang and Q.~Ji, ``Video affective content analysis: A survey of
  state-of-the-art methods,'' \emph{IEEE Transactions on Affective Computing},
  vol.~6, no.~4, pp. 410--430, Oct. 2015.

\bibitem{Croft2015}
W.~B. Croft, D.~Metzler, and T.~Strohman, \emph{{Search engines: Information
  retrieval in practice}}.\hskip 1em plus 0.5em minus 0.4em\relax Pearson
  Education, Inc., 2015, vol. 283.

\bibitem{Gabrilovich2007}
E.~Gabrilovich and S.~Markovitch, ``Computing semantic relatedness using
  wikipedia-based explicit semantic analysis.'' in \emph{IJcAI}, vol.~7, 2007,
  pp. 1606--1611.

\bibitem{Simonetta2018}
F.~Simonetta, ``{Graph based representation of the music symbolic level. A
  music information retrieval application},'' Master's thesis, Università di
  Padova, Apr. 2018.

\bibitem{Mueller2007}
M.~M{ü}ller, ``{Dynamic Time Warping},'' in \emph{Information Retrieval for
  Music and Motion}.\hskip 1em plus 0.5em minus 0.4em\relax Springer Berlin
  Heidelberg, 2007, pp. 69--84.

\bibitem{Bishop2007}
C.~M. Bishop, \emph{Pattern recognition and machine learning, 5th Edition},
  ser. Information science and statistics.\hskip 1em plus 0.5em minus
  0.4em\relax Springer, 2007.

\bibitem{Degara2010}
N.~Degara, A.~Pena, M.~E.~P. Davies, and M.~D. Plumbley, ``Note onset detection
  using rhythmic structure,'' in \emph{2010 IEEE International Conference on
  Acoustics, Speech and Signal Processing}, Mar. 2010, pp. 5526--5529.

\bibitem{MeseguerBrocal2018}
G.~Meseguer-Brocal, A.~Cohen-Hadria, and P.~Geoffroy, ``Dali: A large dataset
  of synchronized audio, lyrics and notes, automatically created using
  teacher-student machine learning paradigm.'' in \emph{19th International
  Society for Music Information Retrieval Conference}, ISMIR, Ed., Sep. 2018.

\bibitem{Maestre2017}
E.~Maestre, P.~Papiotis, M.~Marchini, Q.~Llimona, O.~Mayor, A.~Pérez, and
  M.~M. Wanderley, ``Enriched multimodal representations of music performances:
  Online access and visualization,'' \emph{Ieee Multimedia}, vol.~24, no.~1,
  pp. 24--34, Jan. 2017.

\bibitem{Li2018}
B.~Li, X.~Liu, K.~Dinesh, Z.~Duan, and G.~Sharma, ``Creating a multi-track
  classical music performance dataset for multi-modal music analysis:
  Challenges, insights, and applications,'' \emph{IEEE Transactions on
  Multimedia}, p.~1, 2018.

\bibitem{Pan2010}
S.~J. Pan and Q.~Yang, ``A survey on transfer learning,'' \emph{{IEEE}
  Transactions on Knowledge and Data Engineering}, vol.~22, no.~10, pp.
  1345--1359, oct 2010.

\bibitem{Ntalampiras2017}
S.~Ntalampiras, ``A transfer learning framework for predicting the emotional
  content of generalized sound events,'' \emph{The Journal of the Acoustical
  Society of America}, vol. 141, no.~3, pp. 1694--1701, mar 2017.

\end{thebibliography}
\end{document}